\documentclass[prl,aps,twocolumn,showpacs,nofootinbib,preprintnumbers,amsmath,amssymb,floatfix]{revtex4-1}

\usepackage{graphicx}
\usepackage{dcolumn}
\usepackage{bm}
\usepackage{psboxit,epsfig,wrapfig,boxedminipage,enumerate}
\usepackage{nicefrac}
\usepackage{color}
\usepackage{rotating}

\newcommand{\tb}{}

\newcommand{\vect}[1]{\mathbf{#1}}
\newcommand{\kt}{k_{\rm B}\hat{T}}
\newcommand{\dimT}{T}
\newcommand{\rad}{L}

\begin{document}

\title{{Shock waves in capillary collapse of colloids: a model system
for two--dimensional screened Newtonian gravity}}

\author{J.~Bleibel$^{1,4}$,  S.~Dietrich$^{1,2}$, A.~Dom\'\i nguez$^3$,
  and M.~Oettel$^4$}  
\affiliation{$^1$Max-Planck-Institut f\"ur Intelligente Systeme,
  Heisenbergstr.~3, 70569 Stuttgart, Germany\\
  $^2$Institut f\"ur Theoretische und Angewandte Physik,
  Universit\"at Stuttgart, Pfaffenwaldring 57, 70569 Stuttgart, Germany\\
  $^3$F\'\i sica Te\'orica, Universidad de Sevilla, Apdo.~1065,
  41080 Sevilla, Spain\\
  $^4$Institut f\"ur Physik, WA 331, Johannes Gutenberg
  Universit\"at Mainz, 55099 Mainz, Germany}

\date{\today}

\begin{abstract}
Using Brownian dynamics simulations, density functional theory, and analytical
perturbation theory we study the collapse of a patch of interfacially
trapped, micrometer--sized colloidal particles, driven by long--ranged
capillary attraction. This attraction {is formally analogous} to two--dimensional (2D) screened
Newtonian gravity with the capillary
length $\hat{\lambda}$ as the screening length.
Whereas the limit  $\hat{\lambda} \to \infty$ corresponds to
the global collapse of a self--gravitating fluid, for finite $\hat{\lambda}$
we predict theoretically and observe in simulations
a ringlike density peak at the outer rim of a disclike patch, moving as an
inbound shock wave. 
Possible experimental realizations are discussed.  
\end{abstract}



\maketitle


The dynamics of matter under the influence of long--ranged attractions
is studied intensively in several branches
of physics~\cite{CDR09}, in particular {with respect to inherent} 
instabilities.
Most prominently, the structure formation in the universe is {understood as the 
{consequence} of an instability in self--gravitating matter},
and cosmological theories in
conjunction with numerical simulations
have been successfully applied 
to explain the dynamical
formation of clusters, galaxies, or dark matter halos on large
scales~\cite{Springel:2005}.
More recently, 
other systems with gravitational--like
attractions have been investigated, including seemingly unrelated phenomena
like bacterial chemotaxis \cite{Keller:1970,ChDe10} 
or {capillary--driven clustering in colloids trapped at fluid interfaces}~\cite{Dominguez:2010,Bleibel:2011}. 
{In these systems the interaction is effectively cut off beyond a finite range,
albeit much larger than the mean interparticle separation.}

{In a self--gravitating fluid \textit{any} homogeneous mass
  distribution is unstable with respect to small fluctuations on 
  sufficiently large scales~\cite{Jeans:1902} (\textit{Jeans'
    instability}). In systems with a cut-off gravitational--like
  attraction, this instability only occurs below a critical
  temperature~\cite{Dominguez:2010,ChDe10}. As the range of the
  interaction is scaled~\cite{Bleibel:2011} from infinity down to a
  microscopic length like the size of the particles, the dynamical evolution of the
  instability crosses over from gravitational collapse 
  to spinodal decomposition. 
  A standard theoretical approach to the gravitational collapse in
  cosmology is the so-called cold collapse approximation (see,
  e.g., Ref.~\cite{SaCo95}), within which any force other than gravity (in
  particular the thermal pressure of the fluid) is neglected
  altogether. The view on applications to other physical
  situations raises the natural question how the phenomenology of this
  scenario is affected by a nonvanishing {thermal} pressure and a large
  but finite range {of attractions} and {specifically} how the crossover to the spinodal decomposition
  scenario occurs.}
\begin{figure}[ht]
  \epsfig{file=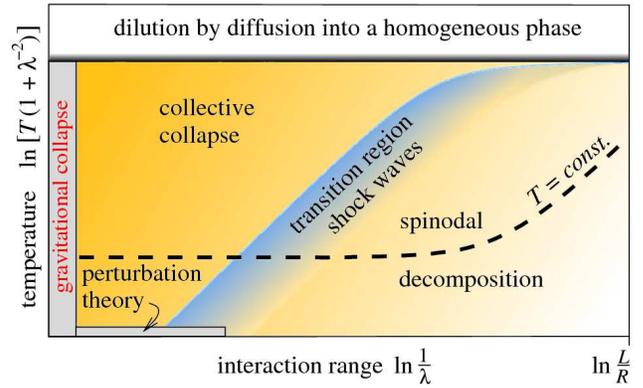,width=0.96\linewidth}
  \caption{\label{fig:phase}Proposed sketch of dynamical regimes for a
    circular patch of radius $\rad$ with particles of radius
    $R$ as function of the range $\lambda = \hat{\lambda}/L$ of the
    interaction and a rescaled effective temperature
    $T(1+\lambda^{-2})$ (see Eq.~(\ref{eq:Teff}) \tb{and below}) 
    The rescaling factor for $T$ leads to a
    horizontal border separating the collapse and dilution regimes. 
    Neglecting a possible temperature dependence of $\hat \lambda$,
    isotherms are parallel to the dashed black line.  {The
      transition region is bounded approximately by the line for which
      linear stability theory for an infinite homogeneous
      distribution~\cite{Dominguez:2010,Bleibel:2011} predicts that
      the fastest growing density mode has a wavenumber $2\pi/L$.}
  }
\end{figure}
\begin{figure*}[ht!]
  \epsfig{file=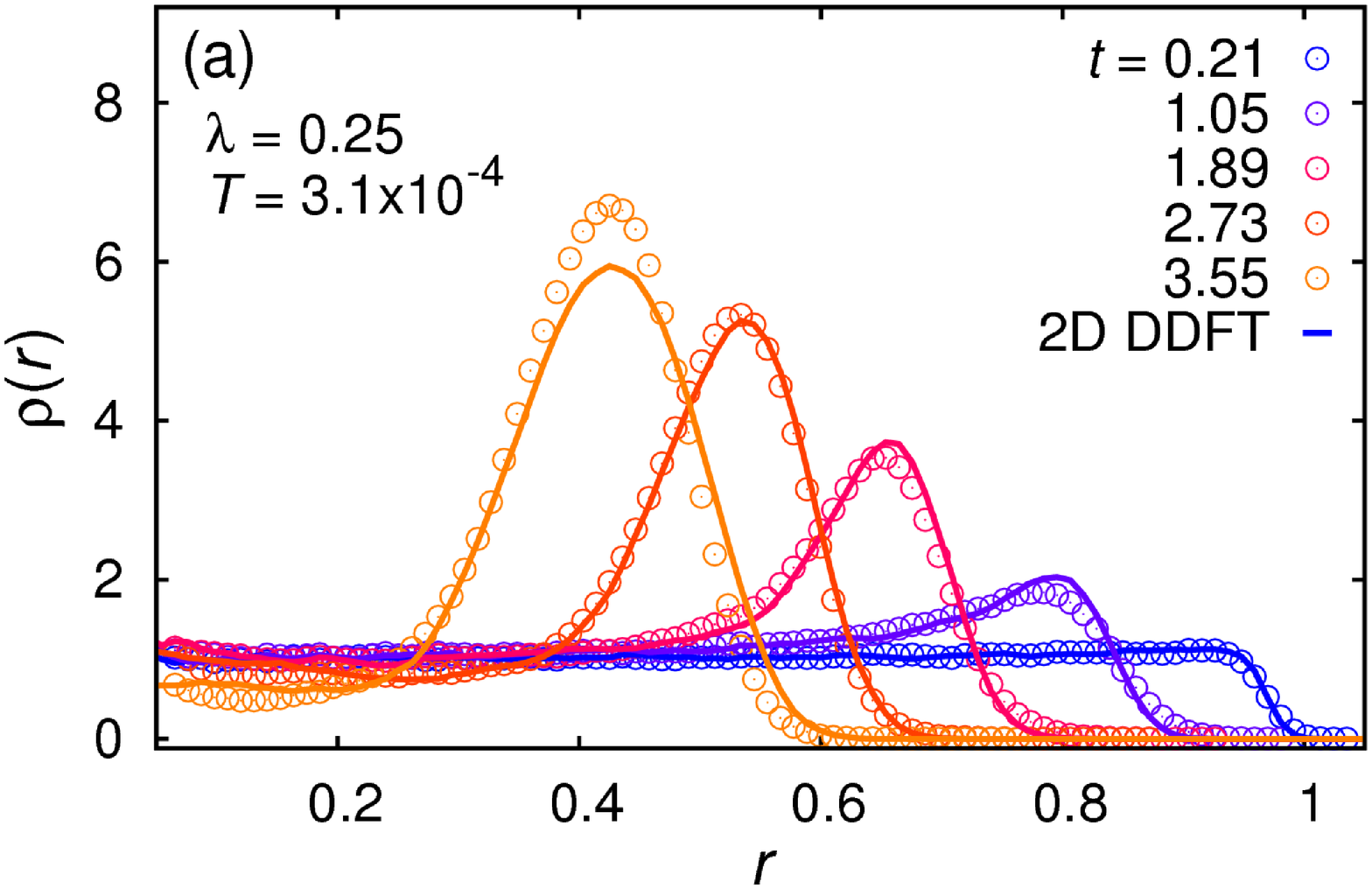,height=8.8cm,angle=270}
  \epsfig{file=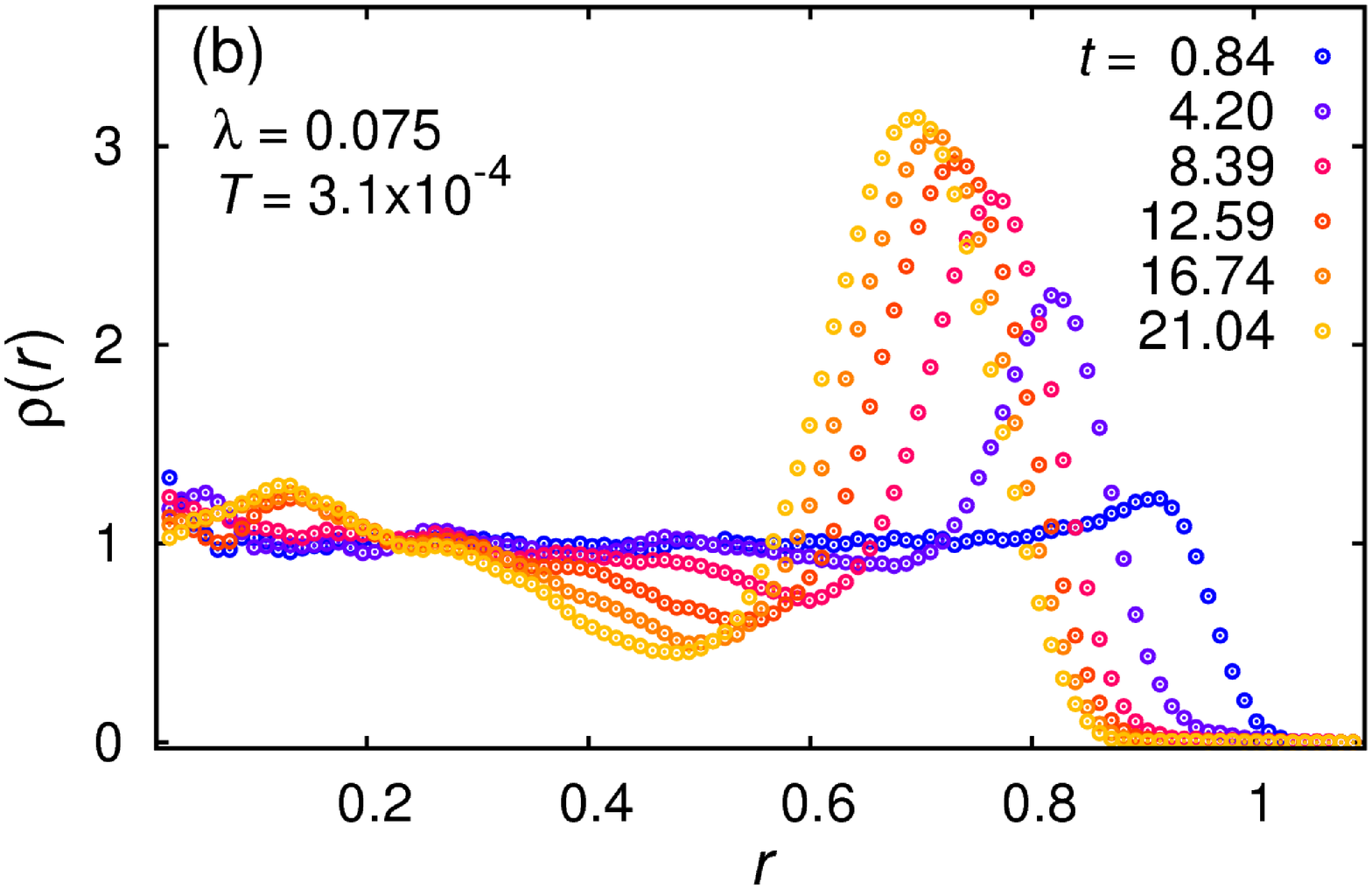,height=8.8cm,angle=270}
  \caption{\label{fig:rho_radial} Evolution of the radial density
    profile for $T=3.1 \times 10^{-4}$ (for further parameters, see the main
    text). 
    Panel (a): $\lambda=0.25$, comparison between 
    2D--DDFT (colored lines), 
    and BD 
    (symbols).  Panel (b): $\lambda=0.075$, only BD
    (DDFT results omitted for
    clarity). {Unlike as in (a), 
    small transient peaks distinct from the one at the outer rim are observed due to} 
    clustering near the center of the
    collapsing disc.} 
\end{figure*} 

Colloidal particles with radii in the micrometer range, which are trapped
at a fluid interface, lend themselves to study this issue. Their weight results in a force $f$ {on each particle}
perpendicular to the interface which deforms it and gives rise
to long--ranged capillary interactions between the particles.
\tb{For large colloid  center--to--center distances $d$, the leading
  interaction term (dominating the collective collapse dynamics) is 
a pair interaction with the potential}~\cite{Oettel:2008}
$V(d)=\left[f^2/(2\pi\gamma)\right] K_0(d/\hat{\lambda})$
with the modified Bessel function $K_0$, the capillary length
$\hat{\lambda}$ ($\sim \mathcal{O}({\rm mm})$), {and the interfacial tension $\gamma$}. For $d < \hat{\lambda}$ {this reduces to 2D Newtonian gravity, $V_0(d) \sim \ln d$},
whereas for $d > \hat{\lambda}$ it
decreases exponentially. This soft matter system is of particular interest because the range
$\hat{\lambda}
=\sqrt{\gamma/(g \Delta\rho_{\rm m})}$ is tunable via the dependence of 
$\gamma$ on temperature $\hat T$ and the concentration of surfactants or 
through the mass density--difference $\Delta\rho_{\rm m}$ between the two
fluid phases forming the interface; 
$g$ is Earth's gravitational acceleration. 

\begin{figure*}[ht]
  \begin{minipage}{0.81\linewidth}
    \begin{center}
      \epsfig{file=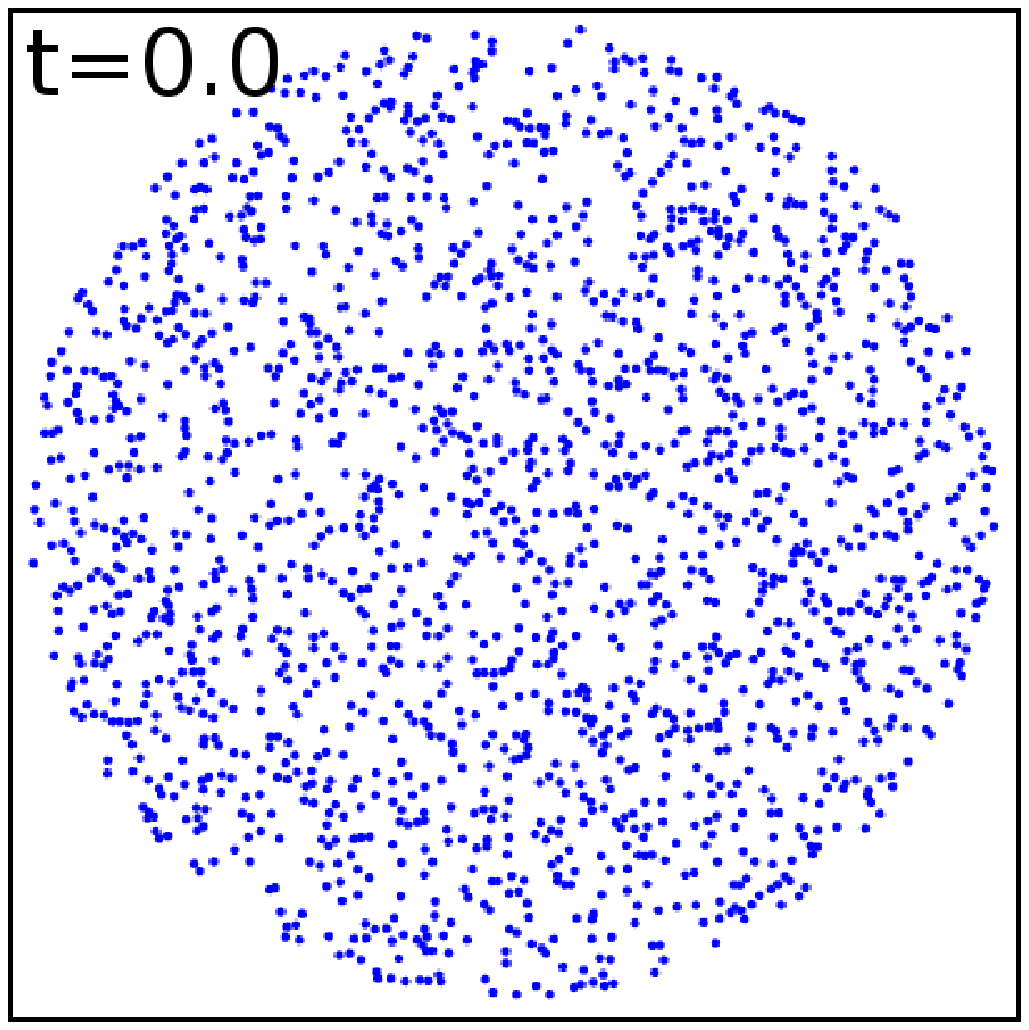,width=.191\linewidth,angle=0}
      \epsfig{file=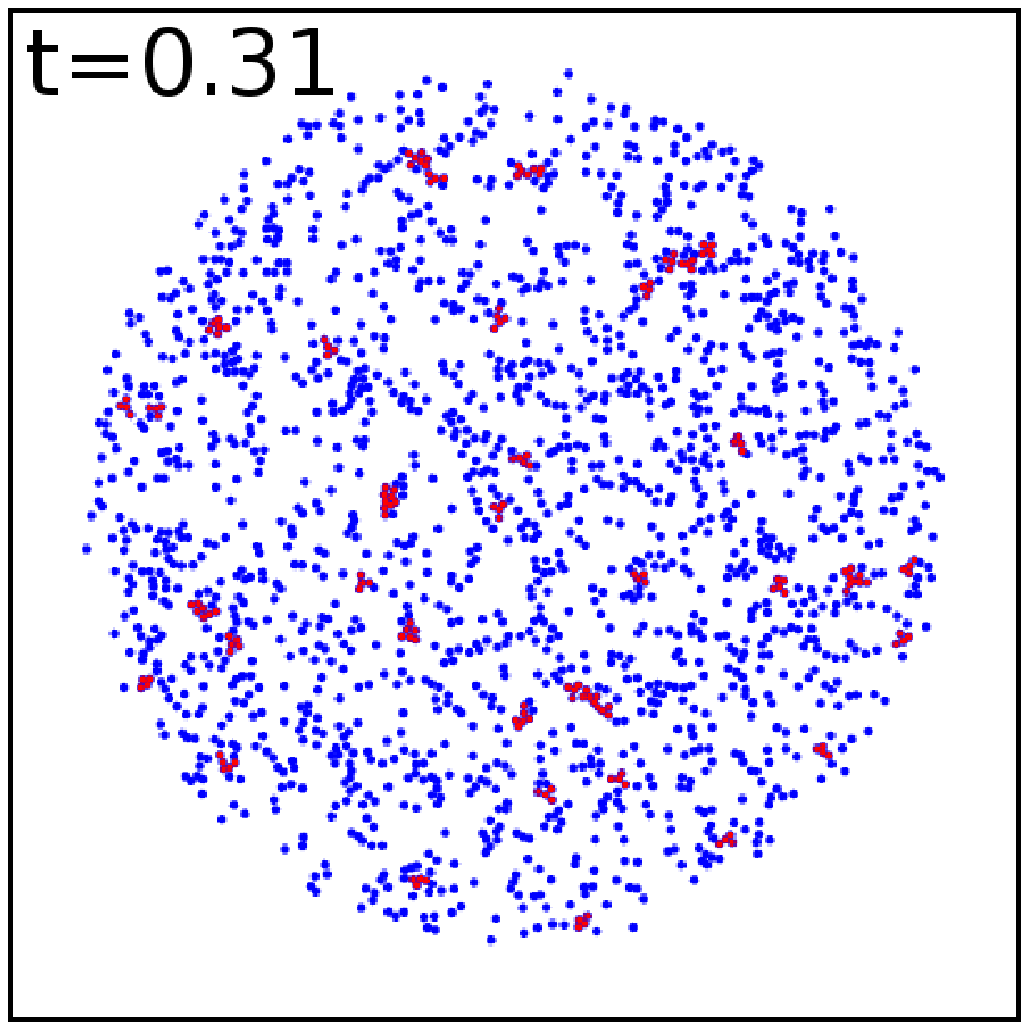,width=.191\linewidth,angle=0}
      \epsfig{file=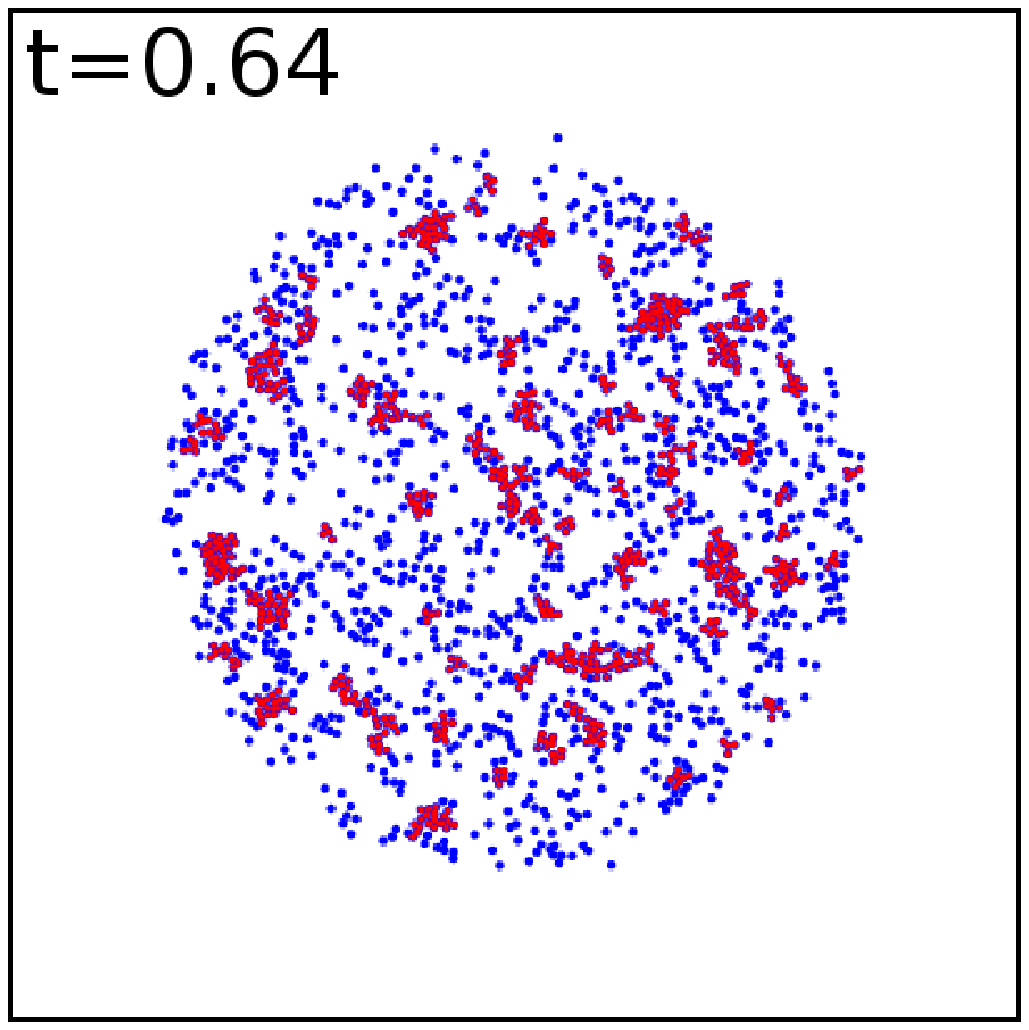,width=.191\linewidth,angle=0}
      \epsfig{file=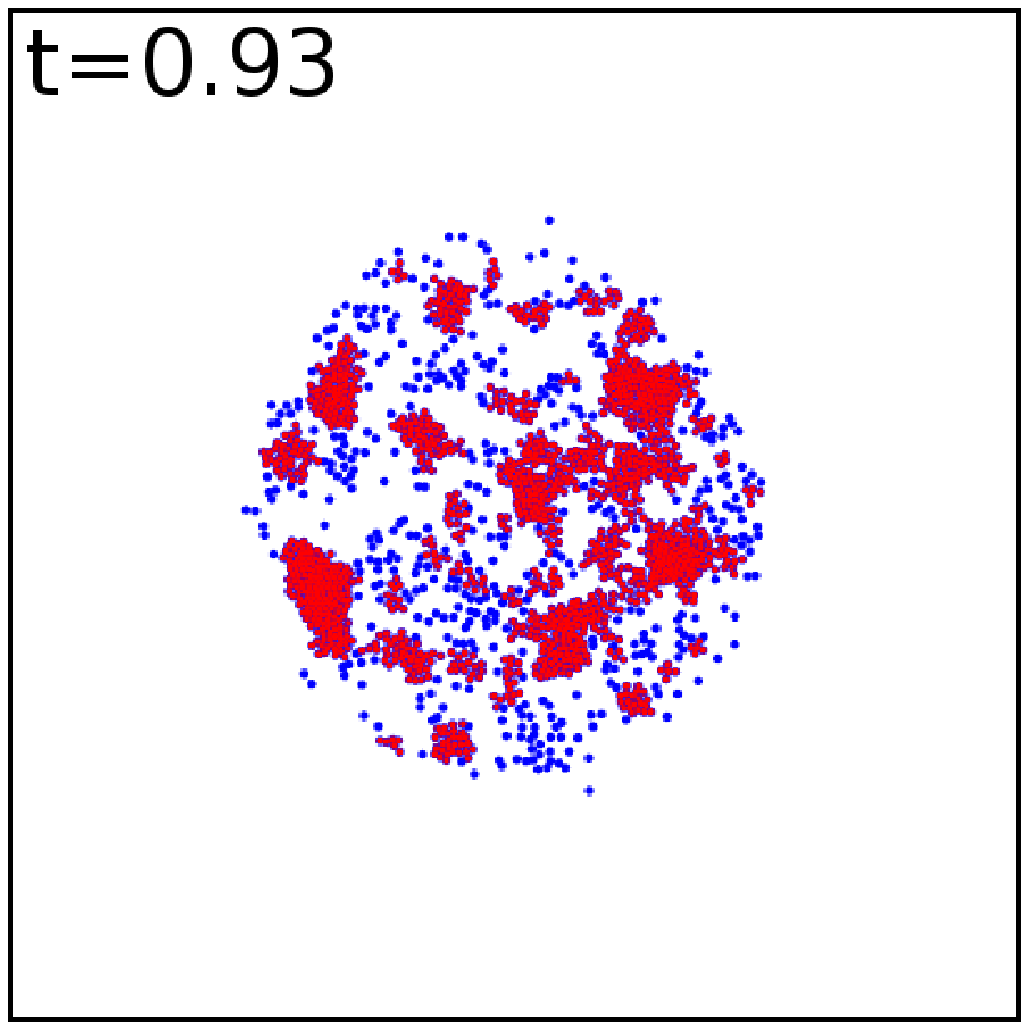,width=.191\linewidth,angle=0}
      \epsfig{file=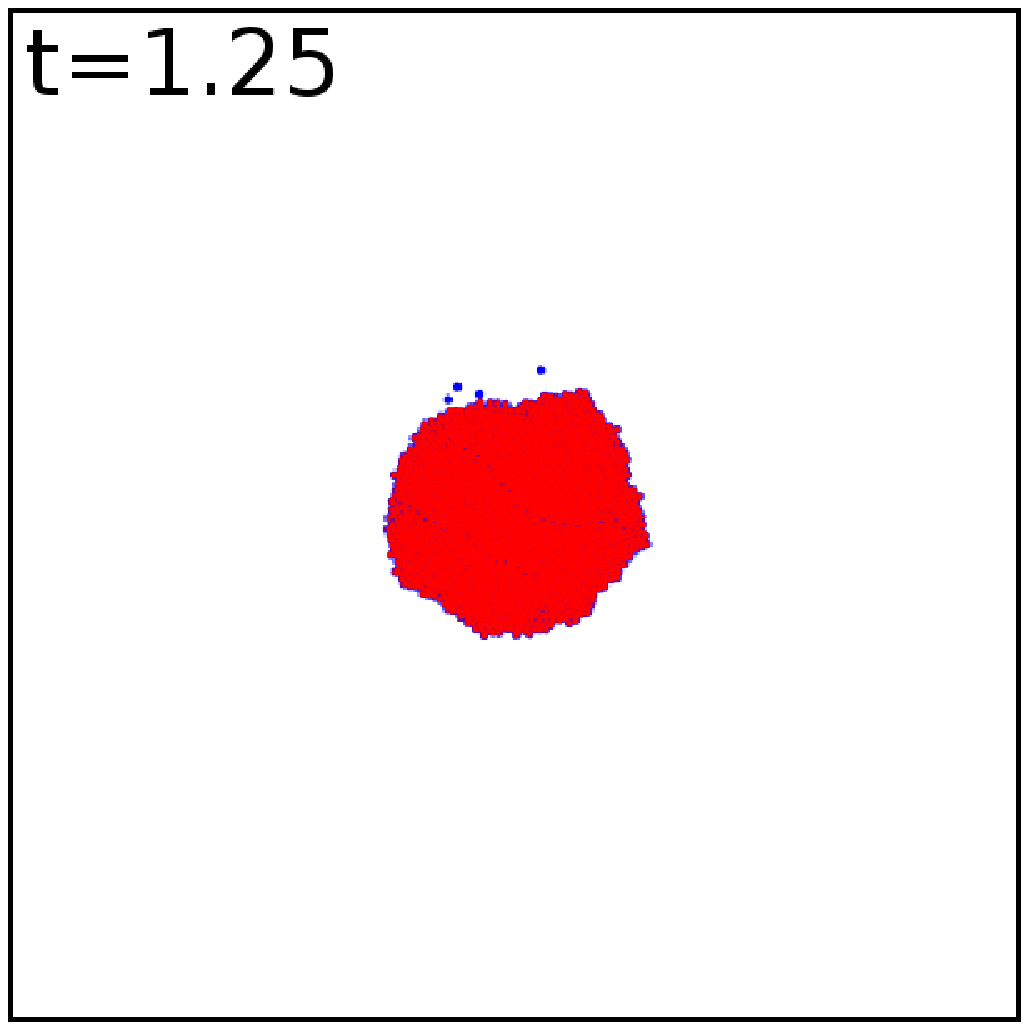,width=.191\linewidth,angle=0}
    \end{center}
  \end{minipage}
  \hspace{0.06\linewidth}
  \begin{minipage}{0.81\linewidth}
    \begin{center}
      \epsfig{file=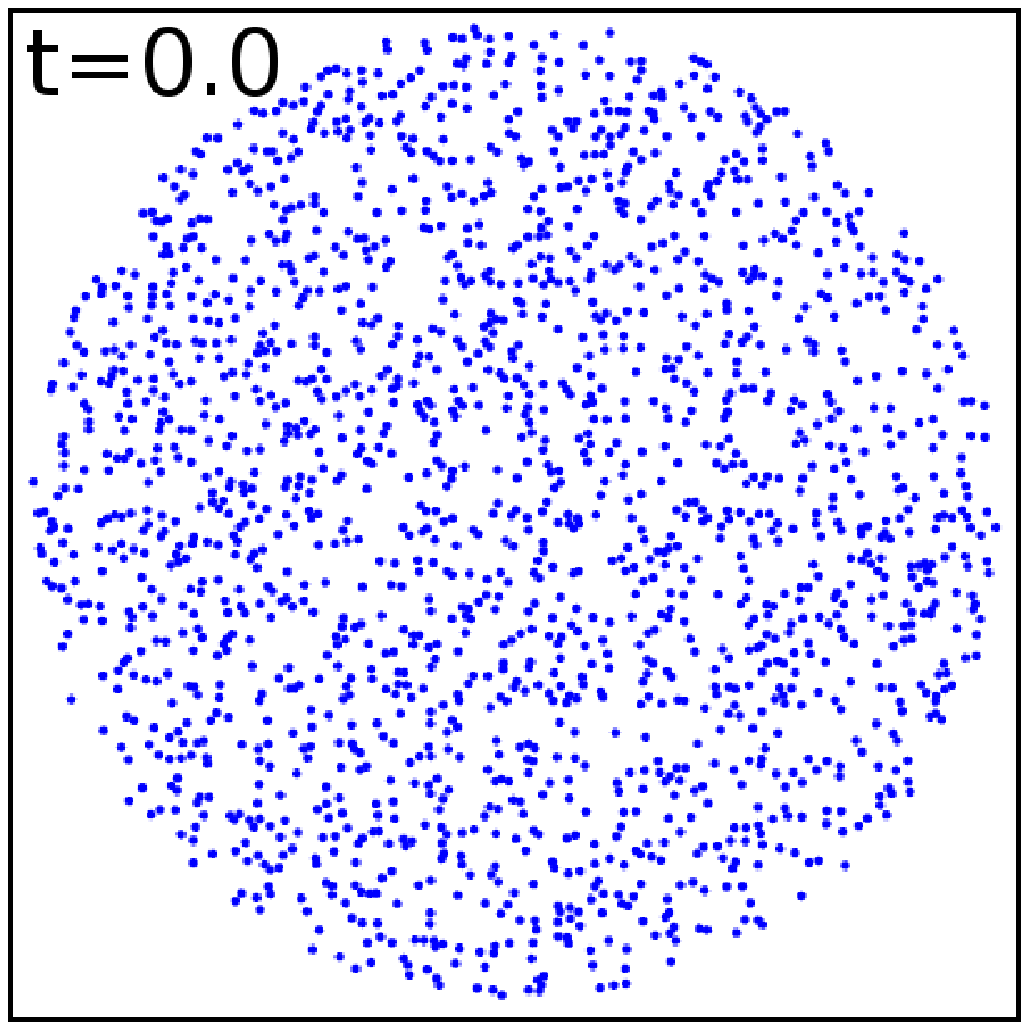,width=.191\linewidth,angle=0}
      \epsfig{file=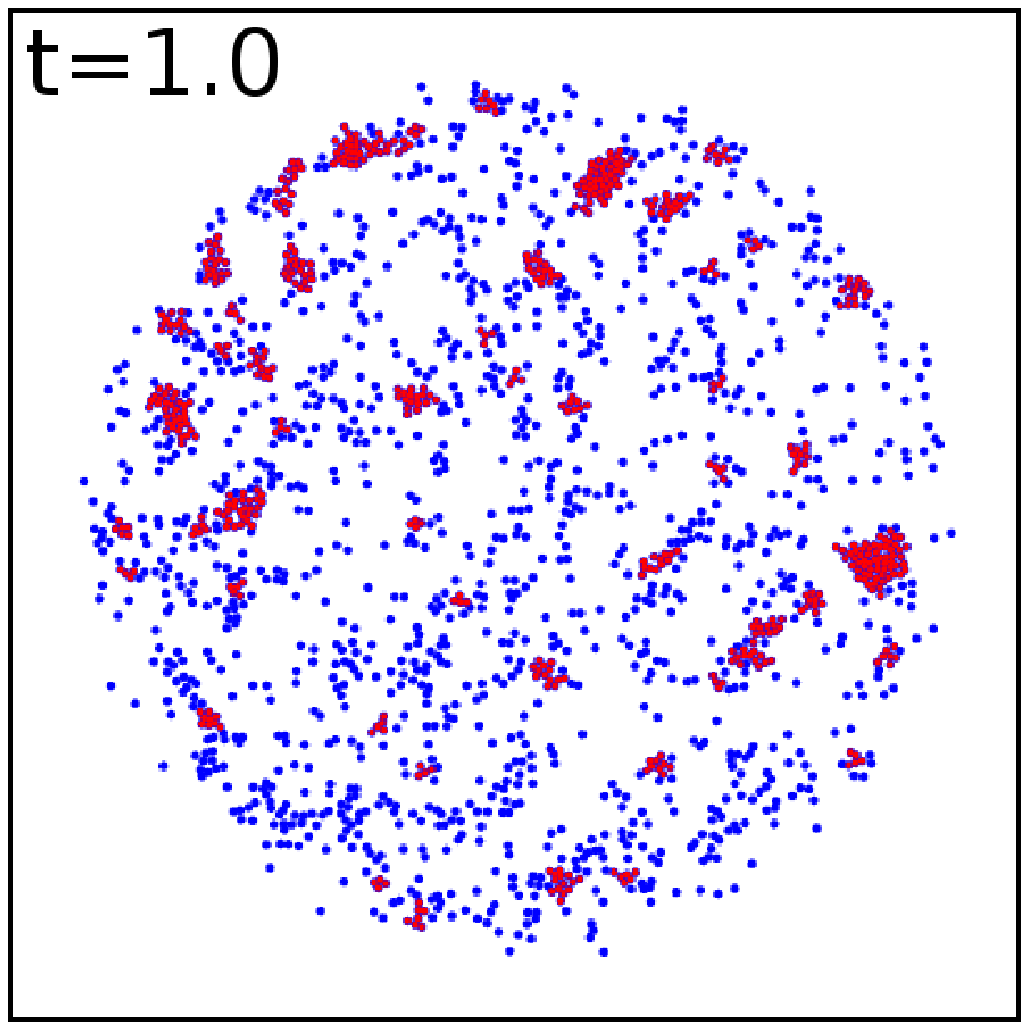,width=.191\linewidth,angle=0}
      \epsfig{file=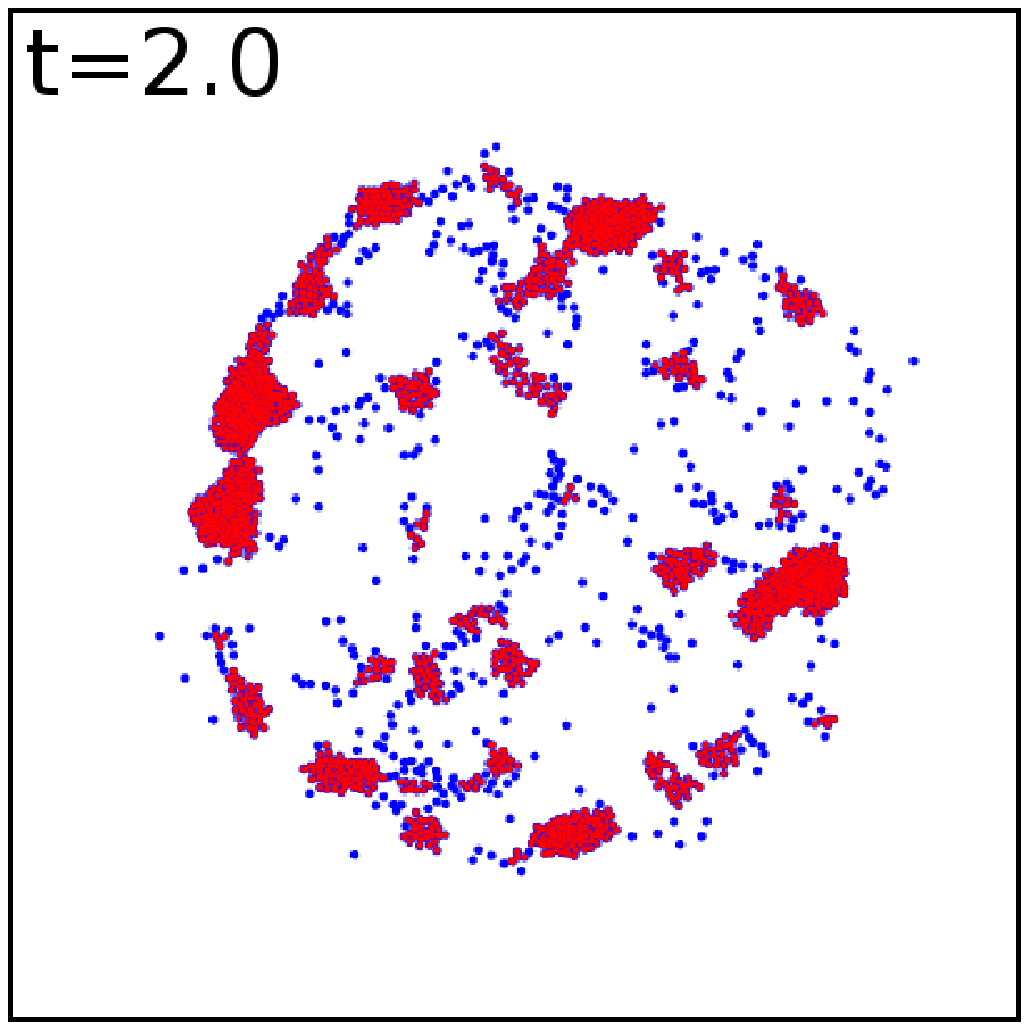,width=.191\linewidth,angle=0}
      \epsfig{file=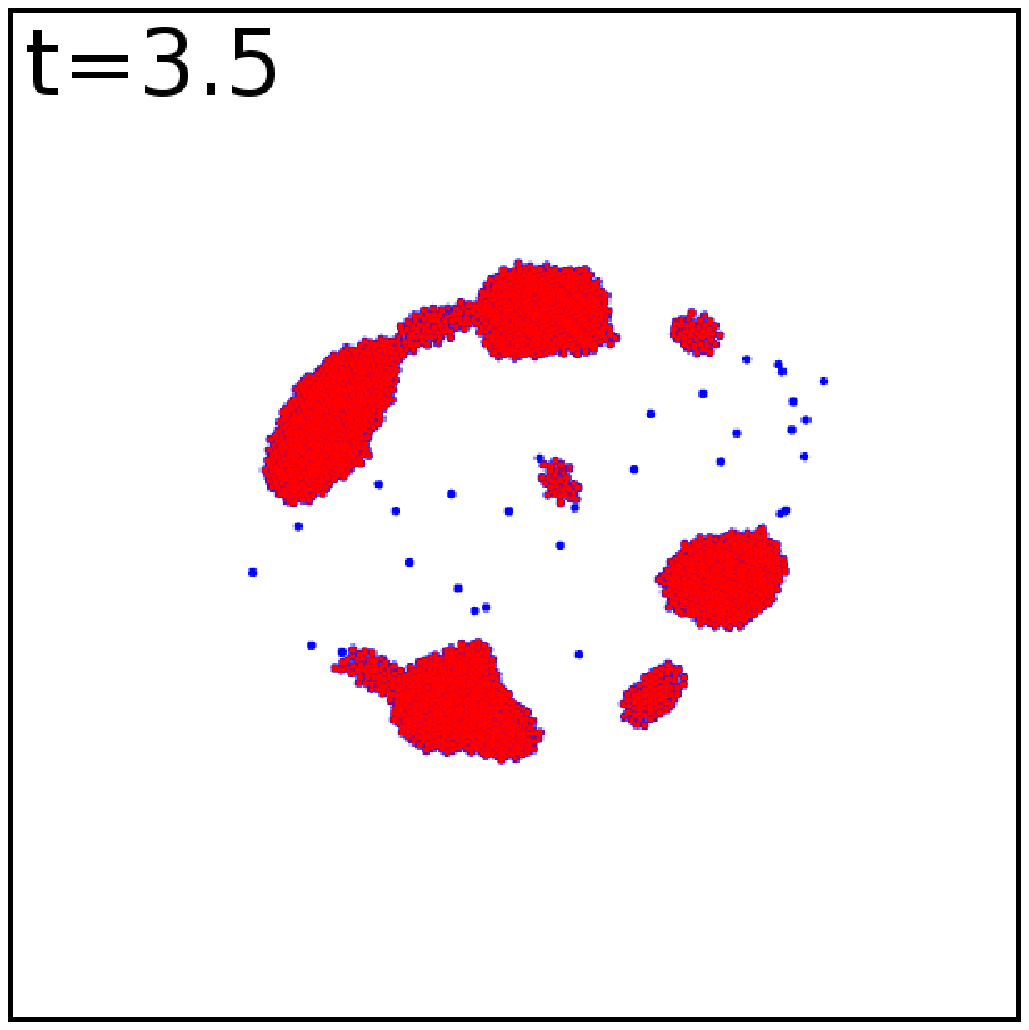,width=.191\linewidth,angle=0}
      \epsfig{file=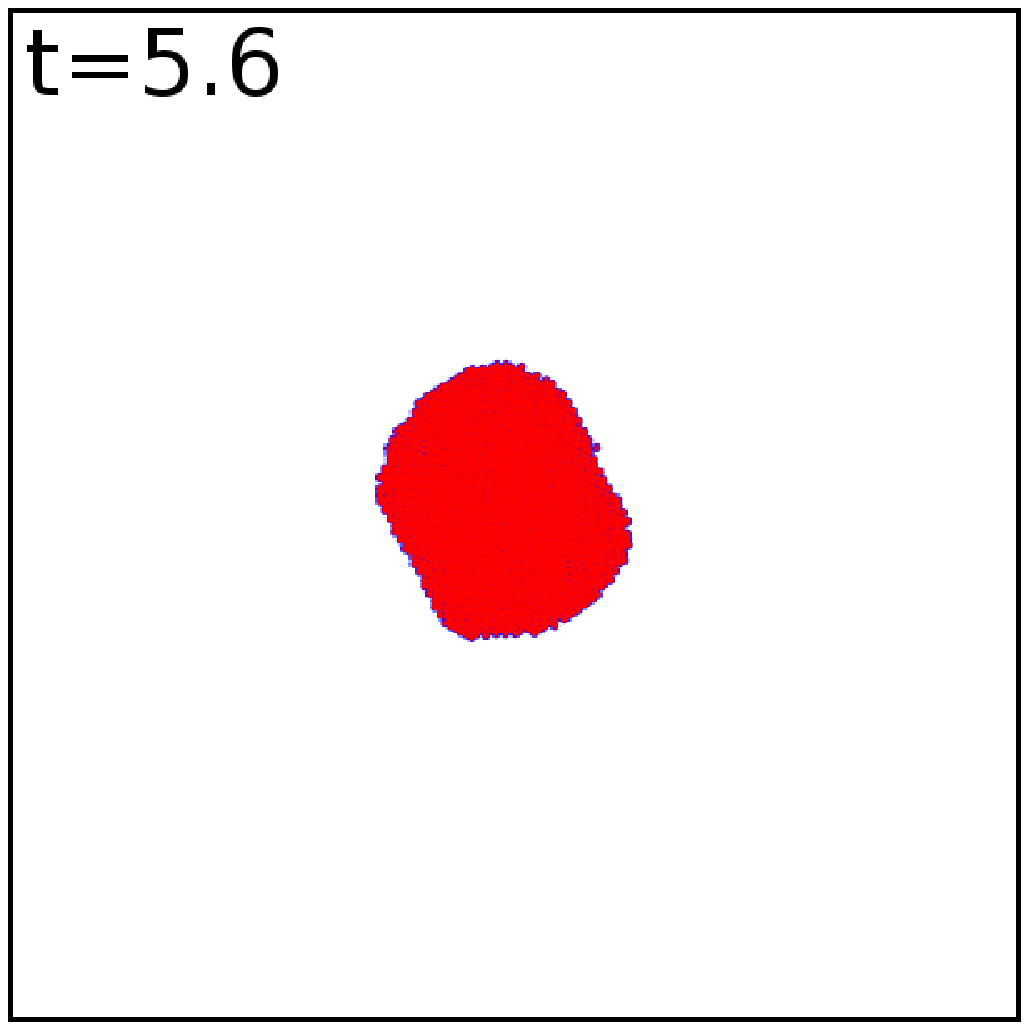,width=.191\linewidth,angle=0}
    \end{center}
  \end{minipage}
  \caption{\label{fig:snaps} Snapshots from BD 
    simulations for temperature $\dimT = 3.1 \times 10^{-4}$ and
    $\lambda=1.5$ (upper row) or $\lambda=0.25$ (lower row).
    Particle distributions in each column have the same radial extent.
    Clusters (i.e., particles with at least 3 neighbors within a distance of
    $3.25 R$) are depicted in red.
    For $\lambda=1.5$ 
    the global collapse appears to be
    faster than the formation of individual smaller clusters.
    For $\lambda=0.25$, small clusters predominantly form at the
    outer rim and collectively move towards the center.
}
\end{figure*}

{\textit{Main Results --} We study the time evolution of the
  2D particle number density $\hat{\varrho}(\hat{\vect
    r}=\{x,y\}, \hat{t})$ of an initially circular patch  of radius
  $\rad$ with {particles of radius $R$} on the flat interface 
  with a homogenous density
  $\hat{\varrho}_0$. 
} {The range is measured by the dimensionless parameter $\lambda =
  \hat{\lambda}/\rad$ and \tb{we introduce} 
  the effective, dimensionless temperature
  \begin{equation}
    \label{eq:Teff}
    T = \frac{\gamma \kt}{f^2 \hat{\varrho}_0 \rad^2}\;. 
  \end{equation}
  \tb{For $\hat\lambda\to\infty$, $T$ is the ratio of the thermal to the
    (mean--field) attractive 
  inner energy of the patch because each particle interacts with
  $N\approx\hat\varrho_0 L^2$ 
  other particles. For $\hat\lambda < L$, the number reduces to 
  $N\approx\hat\varrho_0 \hat \lambda^2$ and this ratio approximately equals
  $(1+\lambda^{-2})T$.}   
  Figure \ref{fig:phase}
  summarizes qualitatively the dynamic phases we have found.
  {If $T$ is large enough (above the black line),
  the} capillary attraction cannot confine the particles to the circular patch,
  {so that it becomes more and more diluted as time progresses}. If $T$ is
  small (dashed line) the patch collapses, but the way how this proceeds depends on
  $\lambda$.  
    In the limit $\lambda\to\infty$ (gravitational collapse),
    the fastest growing modes span the patch and the evolution is
    dominated by the collapse of the structure as a whole.
    In the opposite limit $\lambda\to R$ (spinodal decomposition),
    the fastest growing modes have a characteristic length scale well
    below $\rad$ and the evolution is dominated by the coarsening of
    these domains.
    {We have explored the transition region between these limits} using
    perturbation theory,
    Brownian dynamics (BD) simulations, and dynamic density functional
    theory (DDFT).
    Accordingly, the most prominent
    feature of this transition is a density peak forming at the
    outer rim of the collapsing circular patch which exhibits
    properties of a shock wave. 
     {The latter is defined by the crossing of the characteristic curves
      of a differential equation. For Eq.~(\ref{eq3}) below, in the presence of
      radial symmetry, these characteristics are the trajectories of rings of
      particles.}
    Rings with initially larger radii travel faster than those with
    smaller radii and, upon crossing,  {form a density singularity}
    as $T\to 0$
    (Fig.~\ref{fig:rho_radial}(a)).
    {This feature becomes more pronounced with  $\lambda$ becoming smaller, because the time 
    for building up the shock wave diminishes
    relative to the collapse time of the whole patch
    (Fig.~\ref{fig:snaps}). On the other hand, 
     for smaller $\lambda$ additional small clusters form in the
    interior of the patch and the shock wave amplitude is reduced
    (Fig.~\ref{fig:rho_radial}(b)).}

\textit{Theory -- }
 {We consider a 2D fluid within a mean--field approximation
 appropriate for the long--ranged, capillary interactions and 
 take into account nonzero temperature and short--ranged interactions through 
 a pressure equation of state $\hat{p}(\hat{\varrho}, \hat{T})$
  \cite{Dominguez:2010},} 
 \tb{containing the effects of short--ranged interactions such as hard or soft cores and 
 subleading terms in the capillary forces.}
  The
  particle dynamics is assumed to be in the overdamped regime  with {$\Gamma$ as the associated interfacial
  mobility of the particles} and hydrodynamic interactions
  are neglected.
The relevant time scale of collapse is Jeans' time
$\mathcal{T}=\gamma/(\Gamma f^2 \hat{\varrho}_0)$. We note that the
  time it takes a particle to diffuse a distance $L$ by Brownian
  motion alone is $\sim L^2/(\Gamma \kt) = \mathcal{T}/T$.
Dimensionless variables are introduced as $\varrho=\hat{\varrho}/\hat{\varrho}_0$, $p = \hat{p}/(\kt
\hat\varrho_0)$, $\vect r=\hat{\vect r}/\rad$, 
and $t=\hat{t}/\mathcal{T}$
together with Eq.~(\ref{eq:Teff}). Mass conservation reads~\cite{Dominguez:2010}
\begin{equation}
  \label{eq3}
  \frac{\partial\varrho}{\partial t}=-\nabla\cdot \left(\varrho
    \nabla U[\varrho] - T \nabla p\right),
\end{equation}
with the dimensionless potential of capillary interaction 
$U[\varrho(\vect r)] = 1/(2\pi)\,\int d\vect r^{\prime}\varrho(\vect r^{\prime}) K_0(|\vect r - \vect r^{\prime}|/\lambda)$.
For idealized point particles the pressure is $p_{\rm
  id}=\varrho$; we have also considered a fluid of hard discs of
radius $R$, {described by}~\cite{Grossmann:1997}
$p_{\rm hd}({\varrho})= {\varrho}\,
({\varrho}_c+{\varrho})/({\varrho}_c-{\varrho})$, where
{${\varrho}_c = (2\sqrt{3} \hat{\varrho}_0 R^2)^{-1}$} is the dimensionless density of close packing.}  
Equation~(\ref{eq3}) can be viewed as 
a simple DDFT
for this system \cite{Dominguez:2010} which, 
given the long range of the attractions, is expected to hold at least
for scales much larger than $R$ \tb{and to describe correctly the collective aspects of the dynamics.
Deviations are likely to occur for smaller scales and}
 are mainly attributable to the local density approximation, i.e., the  crude form of the term
$\propto \nabla p$ for the short--ranged \tb{interactions}. Here, {improvement could be achieved} using more
sophisticated expressions from DFT.

The cold collapse (CC) 
scenario ({formally} $\dimT=0$ in Eq.~(\ref{eq3})) is best studied
using {\em La}grangian coordinates: {the characteristic curves {of Eq.~(\ref{eq3})} are the
  radial trajectories of infinitesimally thin rings of particles,
  assuming that the initial radial symmetry is preserved by the
  mean--field evolution. If the initial radius is $r_0$, the
  trajectory is described by a time dependent mapping
  $r=r_{La}(r_0,t)$.}
The Jacobian of this mapping provides the density field in Lagrangian
coordinates:
\begin{equation}
  \label{eq:LagrDensity}
  \frac{1}{\varrho_{La}(r_0,t)} = \frac{r_{La}}{r_0} \frac{\partial r_{La}}{\partial r_0} , 
  \qquad
  (r_0 \leq 1) .
\end{equation}
The mapping is well defined and invertible as long as two rings do not cross. 
If two rings cross, the Jacobian vanishes and the density field
exhibits a singularity.
For $\lambda \to\infty$, the CC approximation 
yields~\cite{Dominguez:2010, Cha11} $r_{La}=r_0\sqrt{1-t}\;$ ($t\leq 1$) and
$\varrho_{La}(r_0,t) = 1/(1-t)$. Thus, {the initial homogeneity
  inside the patch is preserved and a singularity arises at $t=1$
  (i.e., at Jeans' time), when all the rings reach the center
  simultaneously.}
For finite $\lambda$ %
we have applied perturbation theory in terms of $1/\lambda$ leading to
\cite{Bleibel:2011}
\begin{eqnarray}
  \left(\frac{r_{La}}{r_0}\right)^2 & = & 1 - t +
  \frac{2-4\gamma_\mathrm{Euler}-r_0^2 - \ln(4 \lambda^2)}{(4\lambda)^2}t (2-t) 
  - \mbox{} \nonumber \\
  & - & \frac{2 \ln(4 \lambda^2)}{(4\lambda)^2}
  (1-t)^2 \ln |1-t| ,
\end{eqnarray}
\begin{equation}
  \frac{1}{\varrho_{La}(r_0,t)} = \left( \frac{r_{La}}{r_0}\right)^2 - 
   t (2-t) \, \left(\frac{r_0}{4\lambda}\right)^2.
\end{equation}
This result predicts {the formation of an overdensity {for} the
  outermost ring ($r_0=1$)} which becomes {\em singular} ($1/\varrho_{La}=0$)
at a time $t_s \approx 1 + \ln(1.44
\lambda)/(2\lambda)^2$ when it reaches a radius $r_s = r_{La}(1,t_s) \approx 1/(4 \lambda)$. 
{(The singularity is a consequence of the CC approximation; it is
  actually regularized into a density peak moving like a shock wave by the
  term $\propto\nabla p$.) Thus, with decreasing $\lambda$ the collapse singularity occurs later
  and at larger radii.}
Intuitively, for decreasing $\lambda$
the range of the interaction is confined to smaller regions around any
point of the disc. In addition the outer rim of the disc
experiences no balancing pull from the
outside, 
so that there the first overdensity builds up, which then attracts
more and more particles.

\textit{Numerical Methods and Discussion -- }In order to test this theoretical
analysis we have performed simulations and {solved Eq.~(\ref{eq3})
  numerically}.
Simulation parameters were chosen
to reflect the conditions in an actual {experimental
  realization}~\cite{Dominguez:2010} {(patch size $\rad=1.83$ mm,
  particle radius $R=10$ $\mu$m, 
capillary potential depth $f^2/(2\pi\gamma)=0.89\; \kt$, particle mobility $\Gamma=3\pi\eta_{\rm
  water} R$ {with $\eta_\textrm{water}$ being the water viscosity at room
  temperature, and the particle hard--core realized by the repulsive part of
  the Lennard--Jones potential \cite{WCA71}}).}
The simulations were carried out with $N=1804$ particles using
Brownian Dynamics  {(BD)}~\cite{Bleibel:2011,AlTi89} {and the radial density
  profile was {obtained through}  angular and ensemble (120 runs) averages.}
Numerical solutions 
were obtained 
{either through a numerical integration of
  Eq.~(\ref{eq3}) with enforced radial symmetry 
  or through a particle--based
  (Lagrangian) integration scheme of the full 2D equation (labeled as
  2D--DDFT) \cite{Gnedin:2008,Bleibel:2011}.}  
{In this latter case, the density field is probed by a discrete
  number of (fictitious) particles which follow the characteristic
  curves of Eq.~(\ref{eq3}) and the density profile is obtained like
  the profiles from BD simulations.}

{We find that the value} $\lambda=1.5$ (i.e., 
$\hat{\lambda}\approx 2.7$ mm, which is the capillary length of the air--water
interface at ambient conditions) {is at the limit of applicability
  of the perturbation theory}: the DDFT solutions 
for $T=0$ confirm the occurrence of the  singularity at the predicted values
$r_s \approx 0.16$, $t_s \approx 1.1$
and that the formation of the singularity
indeed slows down and occurs at larger values of $r_s$ with decreasing $\lambda$ ($r_s \approx
0.69$, $t_s\approx 1.7$ for $\lambda=0.25$).
The DDFT solutions at nonzero $T=\mathcal{O}(10^{-3}
\dots 10^{-5})$ show that, as for $T=0$, 
an overdensity peak forms at the rim while traveling inwards with increasing amplitude, but the development into a singularity is inhibited.

The results from the BD simulations confirm this 
\tb{and provide further details about the dynamical evolution,}
see Figs.~\ref{fig:rho_radial}~and~\ref{fig:snaps}. 
For $\lambda=0.25$ at $\dimT=3.1 \times 10^{-4}$, 
in contrast to the collapse at larger
$\lambda$, a ring--shaped densified zone forms which is composed of a
number of smaller clusters.
The 2D--DDFT solutions agree well
with the BD simulation data 
for all times. {This indicates that the features we are analyzing
  are not affected by the details of the microscopic
  correlations, the effect of which can be taken into account through
  the macroscopic pressure term.}
Upon decreasing $\lambda$ further, the formation of even more
individual clusters inside the disc is observed, {as a prelude to the
spinodal decomposition scenario}.
For the particular value $\lambda=0.075$ 
the averaged radial density profile still exhibits the collective
behavior of the shock wave at the outer rim {albeit with a smaller amplitude. Additionally}, transient peak
structures for smaller radii 
become also visible and are eventually absorbed by the shock wave traveling inwards.

The phenomenology of local clustering due to small initial fluctuations 
is also visible in the azimuthal direction, both in the BD simulations and
in the 2D--DDFT solution.
In both cases the initial conditions break the radial
  symmetry, {which is expected to be recovered by an ensemble average}. 

Concerning a possible experimental realization, 
a suitable value of $T$ can be arranged using colloids with $R \sim 10$ $\mu$m \cite{Dominguez:2010}.
Observation of the ring--shaped density build-up and the ensuing shock wave requires
$\lambda/\rad \alt 0.25$, corresponding to  {$\rad\gtrsim 10\;\mathrm{mm}$} for an air--water interface.
A likely relevant issue for experiments is the role of
  hydrodynamic interactions.
\tb{
 We have carried out simulations incorporating these on the Rotne--Prager level
(as formulated in Ref.~\cite{Cic04}) for the specific case of particles with contact angle close to zero (i.e., just touching the interface).
Our results 
indicate that the qualitative features of the evolution discussed here are not
affected and the collapse is simply accelerated. This is in line with the
experimentally observed enhanced colloid self--diffusion due to hydrodynamic interactions in such a system
\cite{Zah97}.   
}

\textit{Summary and Conclusions -- } 
We have studied the collapse of a homogeneous, circular patch of colloidal particles
trapped at a fluid interface by means of analytical perturbation theory, Brownian dynamics simulations, 
and dynamical density functional--like
theories. {The capillary attraction is formally analogous to} 
two--dimensional gravity 
with a tunable cutoff length $\lambda$.
We find that a finite value of $\lambda$ strongly influences the collapse features.
While for $\lambda \to \infty$ the {evolution is dominated by the global collapse of the patch,} 
a  {large but} finite $\lambda$ induces the formation of a ring-like overdensity which 
quickly becomes singular {in the limit of a vanishing pressure force (i.e., zero--temperature or cold collapse).} 
A nonvanishing pressure regularizes the singularity into a 
collapsing shock wave. 
{System parameters can be chosen such that
  these spatio--temporal structures can be realized in experiments with micrometer--sized
  colloids.}
Furthermore, this system appears to be {ideally suited} to investigate the
transition from Jeans' gravitational instability
($\lambda \to \infty$) to a spinodal instability ($\lambda \sim $~colloid radius).

\begin{acknowledgments}
\label{acknowl}
{Financial support by the 
  DFG through SFB-TR6 ``Colloids in External Fields'' (Project N01), by DAAD-PPP and by the Spanish
  Government through grants FIS2008-01339 (partially financed by FEDER
  funds) and AIB2010DE-00263 is acknowledged.}
\end{acknowledgments}



\end{document}